\documentclass[notoc]{JHEP}
\usepackage{epsfig}
\usepackage{amsmath}
\usepackage{amssymb}
\usepackage{axodraw}

\def\ra{\rightarrow}

\def\L{{\cal L}}

\def\O{{\cal O}}

\def\qslash{\not{\hbox{\kern-2pt $q$}}}
\def\delslash{\not{\hbox{\kern-2pt $\partial$}}}

\def\beq{\begin{equation}}
\def\eeq{\end{equation}}
\def\eeq{\end{equation}}
\def\bea{\begin{eqnarray}}
\def\eea{\end{eqnarray}}
\def\bq{\begin{quote}}
\def\eq{\end{quote}}

\makeatletter
\def\vereq#1#2{\lower3pt\vbox{\baselineskip1.5pt \lineskip1.5pt
\ialign{$\m@th#1\hfill##\hfil$\crcr#2\crcr\sim\crcr}}}
\makeatother

\title{\center{Supersymmetry Breaking, Fermion Masses
and a Small Extra Dimension}}
\author{David~Elazzar~Kaplan \\ Enrico Fermi Institute, 
University of Chicago, 5640 Ellis Avenue, Chicago, IL 60637 \\
High Energy Physics Division, 
Argonne National Laboratory, Argonne, IL 60439}

\author{Tim~M.~P.~Tait \\ High Energy Physics Division, 
Argonne National Laboratory, Argonne, IL 60439}

\abstract{We present a supersymmetric model in which the observed fermion 
masses and mixings are generated by localizing the three generations
of matter and the two Higgs fields at different locations in a compact 
extra dimension.  Supersymmetry is broken by the shining 
method and the breaking is communicated to standard model fields via
gaugino mediation.  Quark masses, CKM mixing angles and the $\mu$ term
are generated with all dimensionless couplings of $\O(1)$.  All dimensionful
parameters are of order the five-dimensional Planck scale except for the
size of the extra dimension which is of order the GUT scale.  The 
superpartner spectrum is highly predictive and is found to 
have a neutralino LSP over a wide range of parameter space.  The resulting 
phenomenology and interesting extensions of the model are briefly discussed.}
\preprint{hep-ph/0004200 \\ ANL-HEP-PR-00-043 \\ EFI-2000-12}
\keywords{Supersymmetry Breaking, Fermion Masses, Extra Dimensions}

\begin{document}

\section{Introduction}
\label{intro}
\indent \indent
Particle physics is littered with energy scales.  The known particles have 
masses which are spread over numerous orders of magnitude.  
Most hadron masses hover near the scale of presumed quark confinement.
However, the masses of the eight light mesons or pseudo-Goldstone bosons, 
may be parameterized by explicit chiral symmetry breaking in the fundamental 
theory.  In QCD, this breaking for the most part is due to small quark masses.

In the standard model (SM), non-zero quark and charged lepton masses require 
the electroweak symmetry to be broken.  This breaking (EWSB) 
is accomplished by 
allowing a scalar field (Higgs boson) to have a vacuum expectation value (VEV)
thus giving masses to the $W$ and $Z$ bosons, the gauge fields of
the electroweak interaction (minus the photon).  One might expect the 
masses of the charged fermions to be of similar magnitude as the scale 
of electroweak symmetry breaking.  However, instead the masses extend more 
than five orders of magnitude below the weak scale.  One (substantially 
explored) explanation for this large hierarchy of masses is the existence 
of some symmetry broken at high energies producing masses suppressed relative 
to the weak scale.

A more serious hierarchy problem in the SM is the 
instability of the Higgs mass, and thus the electroweak scale, with 
respect to quantum corrections.  A theory with supersymmetry (SUSY) 
softly broken at the weak scale, such as the minimal supersymmetric 
standard model (MSSM), can stabilize the Higgs mass because in such 
theories all quadratically divergent quantum corrections vanish.  
However, generically these theories contain contributions to flavor changing
neutral currents (FCNC) and CP violation which far exceed current 
experimental bounds \cite{GGMS}.  Again, symmetries could restrict flavor 
violations among supersymmetry breaking terms.  For instance flavor 
symmetries which also restrict the soft terms \cite{NS}, gauge symmetries 
which mediate supersymmetry breaking \cite{DN}, or some combination which 
results in partially aligned heavy first two generations \cite{DP,KLMNR}.

Recently, Randall and Sundrum \cite{RS} have suggested a new way to explain
small couplings without appealing to symmetries.  They were able to forbid 
generally flavor violating non-renormalizable operators that mix MSSM fields 
with fields in a supersymmetry breaking sector by spatially separating the 
two sectors in a small extra dimension.  All soft terms appear due to 
contributions coming from the superconformal anomaly \cite{RS,GLMR}.  While 
the soft terms are sufficiently flavor diagonal in the minimal scenario, 
the sleptons are tachyonic, and thus break the electromagnetic
symmetry.  Various model-building scenarios have appeared 
in the literature which attempt to fix this problem \cite{anomfix}.

Arkani-Hamed and Schmaltz have since shown that localizing fields in extra
dimensions at distances of order unity with respect to the fundamental scale
can easily produce exponentially small Yukawa couplings and at the same time
suppress proton decay due to the small overlap the their wave 
functions \cite{AHS}.  This mechanism is especially useful in models in 
which large compact extra dimensions solve the hierarchy problem by bringing 
the fundamental Planck scale down to just above the weak scale \cite{AHDD}.  
In general, this method 
presents an interesting alternative to the usual spontaneous flavor symmetry 
breaking scenarios \cite{flavor}.

In this note, we present a model for fermion masses where Yukawa couplings
are suppressed due to the localization of fields in an extra spatial dimension.
Our model differs significantly from the Arkani-Hamed/Schmaltz model in 
three ways\footnote{Dvali and Shifman \cite{DS2} also considered localizing
complete generations and Higgs fields in an extra dimension and suggested
a way supersymmetry breaking could be included via non-BPS brane 
configurations.  Also, Gherghetta and Pomarol have recently suggested that
small Yukawa couplings could arise from localizing fermions at different
positions in a slice of anti-de Sitter space and discuss variations of
supersymmetry breaking in such scenarios \cite{GP}.}: (i) our extra 
dimension is small (tens of Planck lengths) and supersymmetry solves the 
hierarchy problem, thus avoiding bounds from flavor-violating 
neutral current interactions induced by the relatively light 
Kaluza-Klein (KK) excitations of the gauge bosons \cite{DPQ}.
(ii) Yukawa couplings are small due to the position 
of each generation relative to the (localized) Higgs fields and not due 
to the different splittings of left and right handed fermions \cite{MS}, 
and (iii) our gauge fields fill the entire space in the extra dimension 
thus relieving us of the difficult task of localizing gauge fields in 
field theory \cite{DS1,AHS0}.  

Because our gauge fields live in the bulk, localized supersymmetry breaking 
produces the conditions necessary for recently proposed \cite{KKS,CLNP} 
gaugino-mediated supersymmetry breaking (\~gMSB), arguably the simplest way 
to mediate supersymmetry breaking while avoiding all phenomenological 
flavor constraints.  In addition, by using a particular variation of the 
``shining'' mechanism of Arkani-Hamed, {\it et. al.} \cite{AHHSW}, we 
localize supersymmetry breaking close to the Higgs fields making it possible 
to both generate a $\mu$ term and insure that the right-handed stau is not 
the lightest supersymmetric particle (LSP).  

Our fields are localized by generation and thus are consistent with a 
supersymmetric SU(5) grand unified theory (GUT).  This has implications
for charged lepton masses and thus we shall assume that GUT-breaking is
responsible for producing the correct leptonic spectrum \cite{GJ}.  A 
small extra dimension opens new possibilities with regards to GUT models, 
but we leave this and the leptons for future work.  

The paper is organized as follows.  Section~2 describes all of the elements
necessary for a successful flavor model in this context.  We find the model
to be quite constrained and predictive.  Section~3 describes the incorporation
of supersymmetry and supersymmetry breaking, including a brief review of 
\~gMSB and some attractive modifications with respect to the $\mu$ term.
Section~4 describes the phenomenology of the model with respect to the quarks
and CKM matrix as well as the supersymmetric spectrum.  Section~5 discusses
possible future enhancements of this model and 
an appendix describes the localization of an $N=1$ chiral multiplet .

\section{Flavor from Extra Dimensional Overlaps}
\label{flavor}
\indent \indent
In this section we show how, with a small extra dimension, to generate
small fermion masses from ${\O}(1)$ Yukawa couplings.  The fermions
are localized with respect to the extra dimension and their separations
from each other and the Higgs fields are also ${\O}(1)$ (in Planck units).
Supersymmetry does not play a role in this discussion, other than to
motivate the existence of two Higgs doublets, and thus the results of
this section may be applied generically to any supersymmetric theory.
The details concerning
the localization of a zero mode of a chiral superfield with 
a Gaussian profile
are presented in Appendix \ref{local}.  In this work we will only consider
the quark masses and mixings.  The lepton masses may be obtained from a
straight-forward generalization of these results.

\subsection{Gaussian Localized MSSM fields}
\indent \indent
We take as our starting point a five-dimensional (5d) GUT.
The fifth dimension is small, with associated mass scale
$M_c$ of order $1/100$ times the four-dimensional (4d) Planck scale.
Thus, at energies below $M_c$, there is a 4d effective theory description
of the resulting dynamics.  The fermion masses arise from 5d superpotential
terms, written conveniently in the language of 4d $N=1$ 
supersymmetry \cite{AHHSW} as,
\bea
\int dy \int d^2 \theta \sum^3_{i,j=1} \left\{ \,
  \frac{Y^u_{ij}}{\sqrt{M_*}} H_u \, Q_i \, U^{c}_j
+ \frac{Y^d_{ij}}{\sqrt{M_*}} H_d \, Q_i \, D^{c}_j \,
\right\} + H.c. ,
\label{yukawas}
\eea
where $H_u$ and $H_d$ are the chiral superfields containing the
up- and down-type Higgses, $Q$ is the quark SU(2) doublet,
and $U^c$ and $D^c$ are the up- and down-type quark SU(2) singlets.
The 5d Planck scale $M_*$ is related to the 4d Planck scale
by $M_*=(M_p^2/L)^{1/3}$, $y$ is the coordinate parameterizing the
compact dimension and powers of $M_*$ have been inserted
such that $Y_{ij}$ are dimensionless.  This superpotential
violates the $N=1$ supersymmetry in 5 dimensions 
but is invariant under half of the
supersymmetry transformations which correspond to $N=1$ supersymmetry in 
4 dimensions.
One could imagine explicit breaking in a microscopic theory where at
least some of the fields (either quarks or Higgses)
in (\ref{yukawas}) live on ``3-branes'' which
break translation symmetry in the 5th coordinate along with half of
the supersymmetries\footnote{This assumption could lead to fields with a
simple exponential fall-off rather than the Gaussian profile described
below.  We have checked to see that successful models could be produced
with simple exponentials and found similar results to those described
here.  Thus, for simplicity, we assume all fields have Gaussian wave
functions.}.  We also assume it is possible to localize only a
single left-handed zero-mode, without also introducing a right-handed one.
We will not explore this subtle but difficult issue and assume it can
be accomplished.  Localizing a field using the method outlined in
the appendix also explicitly breaks half the supersymmetries and again
we take this as a requirement of the form of explicit "$N=2$" supersymmetry 
breaking and leave this issue for future work\footnote{We thank Martin 
Schmaltz for discussions on these issues.}.  
In addition, we assume the chiral superfield component of the 5d gauge
multiplet can be given an explicit mass on a 3-brane, or removed from the
low energy theory in some way (for example, see \cite{MP}).
Here we take a
bottom-up approach in constructing the model, allowing successful
phenomenology to motivate the high-energy theory.

The Higgs and fermion fields are localized in the fifth dimension
with Gaussian profiles.  We will assume that the zero modes of the
quarks of a given family are fixed to same
location in $y$, and that the Gaussian profiles of the zero modes
for all three families have a common width $2/\zeta^2$
(which we take to be about the Planck scale).  This is consistent
with GUT scenarios and would arise if all matter was localized
by the same mass function (or VEV) whose slope doesn't significantly
vary over the positions of the localized fields.
The localization of each family around a 
{\em different} point in $y$ can
be accomplished by giving each family's hypermultiplets different
constant mass terms in addition to the single mass profile which
results in the zero modes \cite{AHS}.  Thus, provided the extent of
the extra dimension is large enough that the deviations from this
Gaussian profile are small, the zero mode profile for fermion $j$
is given by,
\bea
\psi^0_j (y) &=& {\left( \frac{2 \zeta_j^2}{\pi}\right)}^{\frac{1}{4}}
\: e^{-\zeta^2_j (y - l_j)^2} .
\eea
The Higgs superfields are also localized in $y$, and we further allow
them to have different widths from the fermions and from each other.
For now we will assume that all of the Yukawa interactions $Y_{ij}$
are exactly 1.  We will see below that a phase (required by CP violation)
will also be important in obtaining the correct mixing angles.

The resulting low energy effective theory has 
exponentially suppressed Yukawa interactions
that result from the overlap of any two fermion wave function with a
Higgs wave function.  We define our coordinate system such that the up-type
Higgs is at $y=0$.  We measure distances in $y$ in units of
$1/\zeta$, and thus the model is completely specified by the locations
of the three families and down-type Higgs, $l_1$, $l_2$, $l_3$, and
$l_h$, and the relative widths of the Higgses,
$r_u = \zeta_{H_u} / \zeta$ and $r_d = \zeta_{H_d} / \zeta$.
In terms of these quantities, the resulting 4d Yukawa
interactions for up-type quarks with $H_u$ is,
\bea
y^u_{ij} &=&
\sqrt{ \frac{r_u}{2 + r_u^2} }
\left( \frac{2^{3/4}}{\pi^{1/4}} \right)
{\rm exp} \left[
{-\frac{1}{2 + r_u^2}
\left( (1 + r_u^2) (l_i^2 + l_j^2) - 2 \, l_i \, l_j \right) }
\right],
\label{overlap}
\eea
where $Y^u_{ij}$ has been taken to be one,
and $\zeta$ to be $M_*$.  There are also
interactions between the down-type quarks and $H_d$
of the same form, but with $r_u \ra r_d$ and
$l_i \ra (l_i - l_h)$.
When the up- and down-type
Higgs scalars obtain VEV's $v_u$ and $v_d$, these interactions will
provide Dirac mass matrices for the quarks.  These matrices may be
diagonalized by separately rotating the right- and left-chiral
fields, resulting in six real quark masses, and the
three mixing angles and one complex phase that make up the
Cabibbo-Kobayashi-Maskawa (CKM) matrix.
Thus, our first task is to see if there exists a suitable arrangement
of parameters to match the low energy data.

\subsection{Fermion Masses and the CKM Matrix}
\indent \indent
We will now outline a method to determine our flavor parameters,
$l_1$, $l_2$, $l_3$, $l_h$, $r_u$, and $r_d$, in
order to fit the low energy data.  We start with the
$\overline{\rm MS}$ quark masses and (90\%
C.L.) three generation CKM matrix\cite{pdb},
\begin{alignat}{3}
\label{data}
m_u({\rm 2~GeV}) &= 1.5 - 5 {\rm~MeV}, &\qquad
m_c(m_c) &= 1.1 - 1.4 {\rm~GeV}, \\
m_d({\rm 2~GeV}) &= 3 - 9   {\rm~MeV}, &\qquad
m_b(m_b) &= 4.1 - 4.4 {\rm~GeV}, \nonumber \\
m_s({\rm 2~GeV}) &= 60 - 170{\rm~MeV}, &\qquad
m_t(m_t) &= 161 - 171 {\rm~GeV}, \nonumber
\end{alignat}
\bea
|V_{CKM}| &=& \left(
\begin{array}{ccc}
0.9742-0.9757 & 0.219-0.226   & 0.002-0.005 \\
0.219-0.225   & 0.9734-0.9749 & 0.037-0.043 \\
0.004-0.014   & 0.035-0.043   & 0.9990-0.9993 \\
\end{array}
\right) \: . \nonumber
\eea
We will work at the top mass scale\footnote{The 
5d Yukawa interactions in Section~2.1 are actually those
at $M_c$.  However, as the quark masses evolve together between $m_t$ and
$M_c$, this can simply be corrected by rescaling all of the 5d
Yukawa interactions.}
(using the three-loop QCD and one-loop
QED renormalization group scaling factors of \cite{scaling} to find the
light quark masses
at $m_t$).

The measured $Z$ boson mass requires $v = \sqrt{v_u^2 + v_d^2} = 246$ 
GeV, while
the ratio $\tan \beta = v_u / v_d$ remains unfixed by EWSB.  We will treat
$\tan \beta$ as a prediction of our model, and characterize the allowed
range of $\tan \beta$ by what results from
the set of model parameters which accurately predict the quark masses and
mixings.  The correct top mass in eq.~(\ref{overlap})
is obtained by fixing the magnitude of $l_3$ once $r_u$ is chosen.
(One can think that $l_3(r_u)$ is fixed by $m_t$ to be a function
of a chosen $r_u$).  In the spirit
of our work, we want to invoke numbers of $\O(1)$, and thus we allow the
widths to vary at most between $1/2$ and 2.
For $v_u \sim v$ and $r_u \sim 1.5$, 
this requires $l_3 \sim 0.3$,
though there is some freedom to vary $l_3$ and $r_u$ together.
Requiring that the ratio $m_t/m_c$ come out
correctly requires us to choose an appropriate $l_2(l_3, r_u) = l_2(r_u)$.
For the particular numbers discussed above the working choice is
$l_2 \sim 2.3$.

Turning to the down-type quarks, the ratio $m_b/m_s$, combined
with the already ``chosen'' value of $r_u$ and the ``determined''
values of $l_3$ and $l_2$ fixes a combination of $r_d$ and
$l_h$.  The magnitude of $m_b$ determines $\tan \beta$ in terms of the
above parameters.
We may now choose $l_1$ in order to arrive at the
correct CKM element $V_{us}$.  As it stands, we have chosen two
parameters ($r_u$ and $r_d$), and used 4 pieces of experimental
data ($m_t$, $m_t/m_c$, $m_b/m_s$, and $V_{us}$) to determine
the remaining parameters $l_1$, $l_2$, $l_3$, and $l_h$.
It remains to determine whether we can accommodate the
remaining experimental data: $V_{ub}$, $V_{cb}$, and the first
family quark masses by varying $r_d$ and $r_u$ independently over the
``reasonable'' range of $1/2$ to 2.  (One can think that two of the
remaining observables fix $r_d$ and $r_u$, and the last two
are predictions of the model).

As it turns out, the answer is no.  While we easily realize the
correct order of magnitude for the remaining predictions, we cannot
quite reach the experimental values.  $V_{cb}$ is always at least
a factor of two or so smaller than its measured value.
This is not really a serious
failing; we have set out to construct a model in which all of the
5d Yukawa interactions were $\O(1)$, but we have tried to realize the
model by taking the couplings to be strictly one.  This will never
produce the CP violation observed in nature \cite{pdb}, and thus
it is obviously too naive.
In fact, taking a simple ansatz that the 5d Yukawa interactions
have the phase structure,
\bea
\label{ansatz}
Y^u = \left(
\begin{array}{rrr}
 1   &  1  &   1  \\
 1   &  1  &   1  \\
 1   &  1  &   1  \\
\end{array}
\right) \qquad
Y^d = \left(
\begin{array}{ccc}
 1   &  i  &   1  \\
 i   &  1  &   i  \\
 1   &  i  &   1  \\
\end{array}
\right) ,
\eea
we find that it is quite easy to fit the low energy data for a
range of $l_h$ and $r_d$.  This particular choice produces
a CP violating phase of $\O(1)$, in accordance with measured
CP violation in the Kaon system.
We have verified that the addition of phases to
the 5d Yukawa interactions greatly extends the workable range
of $l_1$, $l_2$, $l_3$, $l_h$, $r_u$ and $r_d$.  Thus, in our context
the phases are not only critical for CP violation, but also
to get the right magnitude for the CKM elements.
We stress that the ansatz (\ref{ansatz}) is only one of
a wide range of workable solutions, which is a general indication of the
robustness of the result.

\FIGURE[t]{
\epsfysize=4.0in
\centerline{\epsfbox{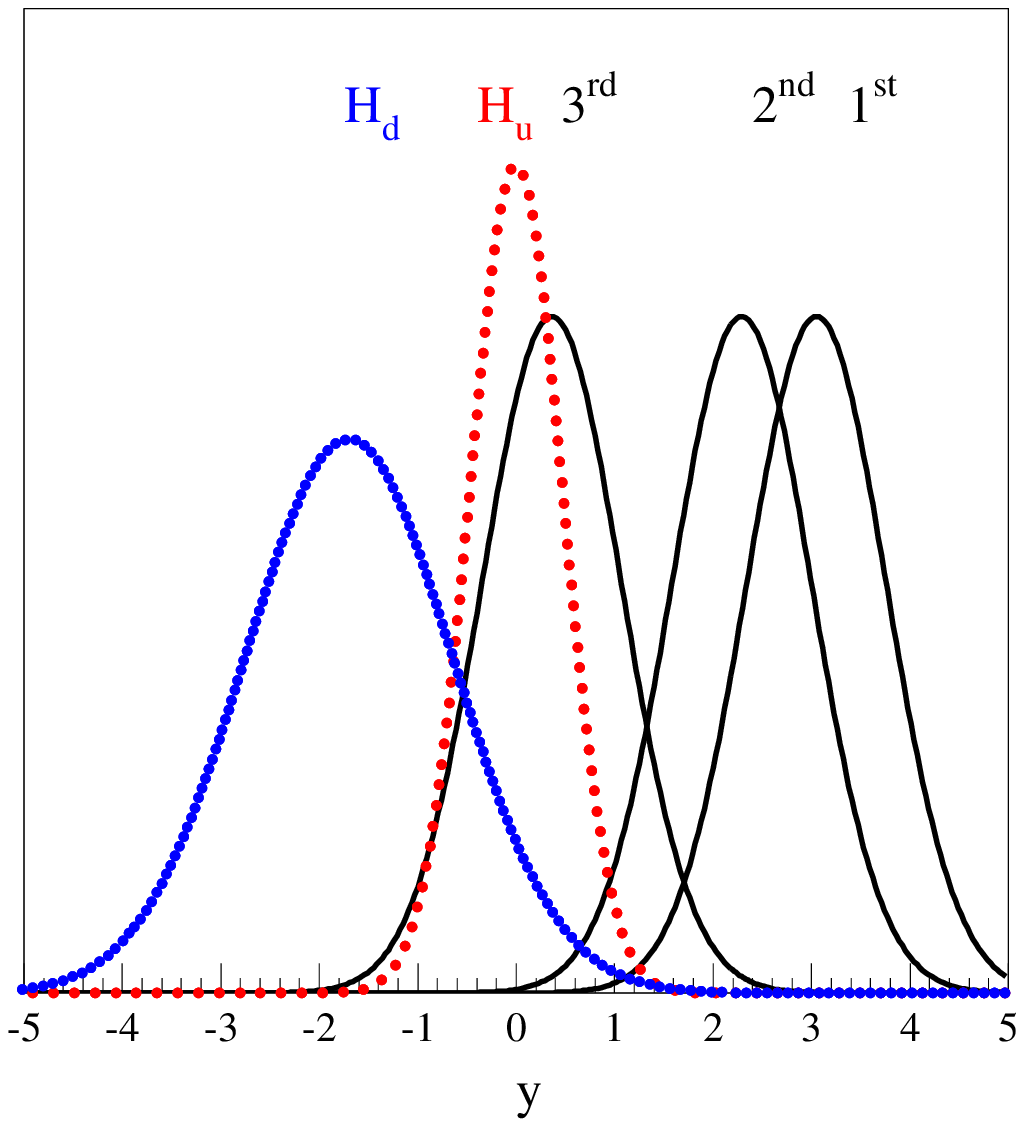}}
\caption{Zero mode profiles for the particular parameter set
with $l_1 = 3.05$, $l_2 = 2.29$, $l_3 = 0.36$, $l_h = -1.7$,
$r_u = 1.5$, and $r_d =0.67$.}
\label{flavfig}}

As a particular example of a working parameter set, consider the
model defined by $r_u = 1.5$, $r_d = 0.67$, $l_1 = 3.05$,
$l_2 = 2.29$, $l_3 = 0.36$, and $l_h = -1.7$.  The Gaussian profiles for
the zero modes are shown graphically in fig.~\ref{flavfig}.  The physical
origin of this solution to the flavor problem is evident from the positions
and widths of the profiles.  The third family sits close to the
up-type Higgs to produce the large value of $m_t$.  The first and
second generations are rather close together compared to the third
generation, in order to correctly generate $V_{us} > V_{cb} > V_{ub}$.
The different widths and locations
of the Higgs profiles account for the fact that $m_t > m_b$ and
$m_c > m_s$, whereas $m_d > m_u$.  The wider down-type Higgs profile
allows it more overlap with the first generation relative to the
narrow up-type Higgs profile, which is very small at the distant
location of the first family.
This results in low energy predictions,
\begin{alignat}{3}
\label{predict}
m_u({\rm 2~GeV}) &= 3.6{\rm~MeV}, &\qquad 
m_c(m_c) &= 1.3{\rm~GeV}, \\
m_d({\rm 2~GeV}) &= 8.8{\rm~MeV}, &\qquad 
m_b(m_b) &= 4.3 {\rm~GeV}, \nonumber \\
m_s({\rm 2~GeV}) &= 75.0{\rm~MeV}, &\qquad
m_t(m_t) &= 166 {\rm~GeV}, \nonumber
\end{alignat}
\bea
|V_{CKM}| &=& \left(
\begin{array}{lll}
   0.9755 & 0.221  & 0.0046 \\
   0.220  & 0.9748 & 0.038  \\
   0.0076 & 0.037  & 0.9992 \\
\end{array}
\right) \: . \nonumber
\eea
Comparison of eqs.(\ref{data}) and (\ref{predict}) indicates that we
have succeeded admirably 
in satisfying the low energy flavor measurements.  The CKM elements
fit comfortably into their allowed ranges, and the quark masses are
all reasonable.  Furthermore, the quantities 
$(m_u + m_d)/2 = 6$~MeV,
and $m_u / m_d = 0.4$ are more rigorously defined
from Chiral Perturbation Theory (CPT), and fall 
into the acceptable ranges \cite{pdb}.

This particular model has $\tan \beta = 13$.  In fact there is some
remaining freedom to move around in the acceptable ranges for the
quark masses and mixings, provided all of the parameters are
suitably adjusted together.  This can be characterized as $l_h$,
which shows the greatest freedom to move consistently with the data.
In Fig.~\ref{tblhfig} we show a scatter plot of models consistent with
low energy data, in the plane of $l_h$ and the resulting $\tan \beta$.
As can be seen from the figure, $l_h$ can vary between about
-0.5 and -2.4, with $\tan \beta$ running from 45 to about 8 in that
range.  The cut-off at $l_h = -2.4$ occurs because going lower
requires $r_d < 1/2$.

\FIGURE[t]{
\epsfysize=3.5in
\centerline{\epsfbox{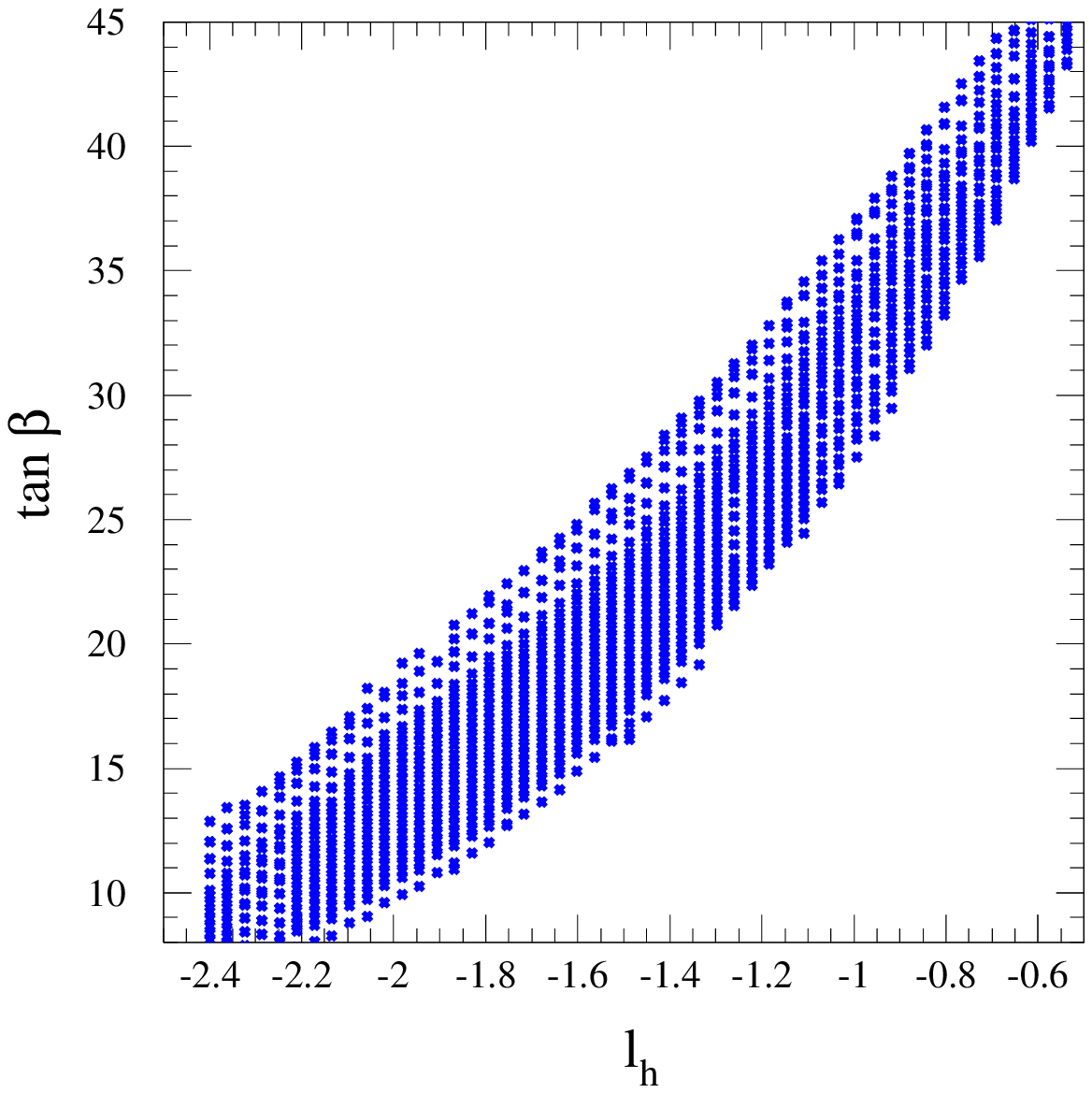}}
\caption{Scatter plot for models with various $l_h$ and the resulting
$\tan \beta$.}
\label{tblhfig}}

\section{Supersymmetry Breaking}
\label{susyb}
\indent \indent
In this section we introduce supersymmetry breaking into the model described
above.  Since our model involves a small extra dimension with SM gauge
fields in the bulk, it naturally lends itself to the mechanism of 
gaugino-mediated supersymmetry breaking \cite{KKS,CLNP}.  If we use the 
shining mechanism to break supersymmetry \cite{AHHSW}, we can naturally 
produce a weak-scale $\mu$ term still keeping all coefficients of ${\O}(1)$.  
We shall see that a general consequence of this $\mu$ term solution is a 
non-zero soft mass for the down-type Higgs at the high scale.  This results
in a neutralino LSP in a large region of parameter space and thus we avoid 
bounds on stable charged particles.

\subsection{Gaugino Mediated Supersymmetry Breaking}
\indent \indent
It was realized a number of years ago that a very simple set 
of boundary conditions on soft terms, namely a non-zero gaugino mass at a high
scale, would produce a theory at the weak scale with scalar masses and gaugino
masses of the same size, suppressed contributions to FCNC and successful 
EWSB.  Models with these boundary conditions were
called ``no scale'' models \cite{noscale}.  Gaugino mediation \cite{KKS,CLNP} 
gives (so far) the simplest realization of this spectrum from a microscopic 
theory.

For \~gMSB to work, we need a small ($\sim 10 - 100$ Planck lengths) extra 
dimension in which the SM gauge fields and their superpartners 
propagate.  We also require supersymmetry to break on a 4d hypersurface 
which is separated in the fifth dimension from the hypersurface(s) on 
which MSSM matter lives.  Thus contributions to scalar masses via 
Planck-suppressed operators are also exponentially suppressed as Yukawa 
couplings are in the previous section. 

The effective 4d operator which contributes to gaugino masses is
\begin{equation}
\label{gauginos}
\int d^2 \theta \, \frac{S}{M_* (M_* L)} \, W^{\alpha} W_{\alpha},
\end{equation}
where $L$ is the size of the extra dimension, $W^{\alpha}$ is a chiral 
superfield whose lowest component is the gaugino and $S$ is a gauge singlet.
The field $S$ lives in the supersymmetry breaking sector and has a non-zero 
VEV in its auxiliary component, $F_S$.  As a result, this term produces a 
gaugino mass at the compactification scale of order 
$M_{1/2}\sim \frac{F_S}{M_* (M_* L)}$.

The operators which contribute to squared scalar masses are
\begin{equation}
\label{scalars}
\int d^4 \theta \, \frac{S^{\dagger}S}{M_*^2 (M_* L)} \; Q_i^{\dagger} \, Q_j,
\end{equation}
where $Q_i$ are chiral superfields which contain MSSM matter and $i,j$ are
flavor indices.  The operators with $i\neq j$ would normally make dangerous
contributions to FCNC.  However, all of the terms in (\ref{scalars}) are
multiplied by exponentially suppressed coefficients due to the spatial
separation between the supersymmetry breaking sector and MSSM matter 
fields, and thus these terms may be ignored provided the supersymmetry breaking
sector is sufficiently distant from all MSSM matter.  This
suppression was not operative in the gaugino mass operators because
the gauginos are not localized in $y$.

The dominant contribution to scalar masses comes from the renormalization 
group (RG) evolution\footnote{Our RG analysis is carried out at 
two loops with respect to $\alpha_S$ (with one loop thresholds)
and at one loop with respect to all other quantities using the beta
functions of \cite{MV}.} from the compactification scale 
$M_c\equiv 1/L$ to the weak scale.  The relevant term in the one-loop 
beta function is
\begin{equation}
\frac{d}{dt}m_{\tilde f}^2 = - \sum_i \, 
\frac{g_i^2}{2 \pi^2} \, C_i(r_f) \, M_i^2
\end{equation}
where $m_{\tilde f}^2$ are the squared scalar masses, $M_i$ are the gaugino 
masses and $C_i(r_f)$ is the quadratic casimir for representation $r_f$ of 
chiral superfield $f$ in gauge group $i$.  As in gauge mediation,
these contributions are only proportional to gauge couplings,
and therefor are 
flavor diagonal and do not contribute to FCNC or CP violation.  If we take 
$M_c\sim M_{GUT}$, the loop factor suppression is matched by a large 
logarithm such that scalar masses and gaugino masses are comparable.

\FIGURE[!t]{
\begin{picture}(420,190)(0,0)
%
%
   \Line( 100, 110 )( 145, 185 )
   \Line( 100,  15 )( 145,  90 )
   \Line( 100,  15 )( 100, 110 )
   \Line( 145,  90 )( 145, 185 )
   \Text( 100,   5 )[c]{$y=l_i$}
   \Text(  75, 158 )[l]{localized}
   \Text(  75, 145 )[l]{matter}
   \Text(  75, 132 )[l]{field}
   \DashLine(  137, 155 )( 130, 114 ){4}
   \CArc(      190,  10 )( 120, 60, 120 )
   \PhotonArc( 190,  10 )( 120, 60, 120 ){3}{10}
   \Line(      130, 114 )( 120, 77 )
   \DashLine(  108,  45 )( 120, 77 ){4}
   \CArc(      180, 180 )( 120, 240, 300 )
   \PhotonArc( 180, 180 )( 120, 240, 300 ){3}{10}
   \CArc(      238,  97 )(  21, 278, 412 )
   \PhotonArc( 238,  97 )(  21, 278, 412 ){3}{5}
   \put( 251, 114 ){\circle*{6}}
   \put( 240,  76 ){\circle*{6}}
   \Line( 225, 110 )( 270, 185 )
   \Line( 225,  15 )( 270,  90 )
   \Line( 225,  15 )( 225, 110 )
   \Line( 270,  90 )( 270, 185 )
   \Text( 225,   5 )[c]{$y=l_x$}
   \Text( 270,  68 )[l]{SUSY-}
   \Text( 270,  53 )[l]{breaking}
   \Text( 270,  40 )[l]{singlet}
\end{picture}
\caption{Loop contribution to scalar masses from localized supersymmetry 
breaking.}
\label{loop}}

There is also a contribution to the scalar masses which dominates at the 
compactification scale.  It can be considered a threshold correction
from integrating out the higher KK modes of the gaugino.
It comes from the loop contribution depicted in 
fig.~\ref{loop}, where gauginos run in the loop and the operators 
responsible for gaugino masses are inserted on the propagators at the 
location of supersymmetry breaking \cite{KKS,CLNP}.  Contributions like
this one were calculated in \cite{KKS} assuming that the size of the 
extra dimension and the distance between supersymmetry breaking and
MSSM matter are the same (see also \cite{MP}).  The latter will not be the 
case in our model and these contributions can be important as we discuss
below and in Section~\ref{pheno}.

One remaining superpartner mass which we have not yet specified is that 
of the Higgsino.  The $\mu$ term, which is a superpotential mass term 
mixing the up- and down-type Higgses ($H_u$ and $H_d$), should be 
generated dynamically to explain its weak scale value required by 
radiative EWSB.  By putting the Higgs fields in the bulk 
\cite{CLP,CLNP}, the $\mu$ term can be produced via the 
Giudice-Masiero mechanism \cite{GM}.  It was also pointed out \cite{SS} 
that the shining mechanism \cite{AHHSW}, which could break 
supersymmetry on a distant hypersurface, could also be used to produce 
a $\mu$ term on our brane.  Both solutions require somewhat small 
couplings, the former for soft masses in the Higgs sector and the 
latter for the $\mu$ term itself.

In the standard picture of \~gMSB, with $M_c = M_{GUT}$ and Higgs 
fields localized with MSSM matter, an electrically charged stau is the 
LSP, a scenario which is disfavored by cosmological considerations.  
The lightest neutralino can become the LSP if either the Higgs fields 
are in the bulk and the soft mass of $H_d$ is somewhat larger than the 
soft mass of $H_u$ \cite{CLNP} or $M_c>M_{GUT}$ by at least an order of 
magnitude so that the scalar masses run significantly above the GUT 
scale \cite{SS}.

Here we offer modified solutions to the $\mu$ and LSP problems which naturally
fit into our model of flavor.  The mechanism allows all couplings to
be of $\O(1)$ and produces simple high-scale boundary conditions
(though the  location of supersymmetry breaking is somewhat fine-tuned).  
The weak scale spectrum is in general distinguishable from other 
models of \~gMSB.

\subsection{Soft Parameters and the $\mu$ term}
\label{ourmodel}
\indent \indent
To break supersymmetry in our model, we will use the shining mechanism of
Arkani-Hamed, {\it et. al.} \cite{AHHSW}.  One advantage of shining is that
it does not require the localization of gauge fields, a difficult task in 
more than four dimensions \cite{DS1,AHS0}.  Shining is also advantageous as
it allows for new solutions to the $\mu$ problem \cite{AHHSW,SS}.  We discuss 
two variations of the solution of Schmaltz and Skiba in minimal gaugino
mediation \cite{SS} and show in our context how to remove the requirement 
for a small coupling.  We also show that in this context that a neutralino
can be easily be the LSP even if $M_c = M_{GUT}$.

\subsubsection{Scenario 1}
We introduce two chiral superfields, $\Phi$ and $\Phi^c$, to the bulk 
which are singlets under the SM gauge groups.  These fields together
are a hypermultiplet of the $N=1$ supersymmetry in 5d, which is conserved by this
mechanism up to explicit breaking by a source $J^c$ which couples to $\Phi$ 
in the superpotential.  This source is localized at $y = l_s$
near the MSSM fields.  In the absence of further ingredients, the scalar 
component of $\Phi^c$ will acquire a VEV and supersymmetry remains 
unbroken.  supersymmetry 
is broken by introducing another singlet chiral superfield 
$X$ localized in the bulk far from our MSSM matter, 
and coupling to $\Phi^c$.  
Assuming $X$ has a profile $\psi_x^0(y)$ with respect to 
the compact dimension, the Lagrange density can be expressed,
\bea
\L&=&\int d^4 \theta \int dy \left[ \Phi(y)\Phi^{\dagger}(y) 
        + \Phi^c(y)\Phi^{c\dagger}(y) + \left| \psi_x^0(y) 
                \right|^2 X X^{\dagger}
        \right] \nonumber\\
  & & + \int d^2 \theta \int dy \left[\Phi^c(y)(\partial_y + m_{\phi})\Phi(y) 
        + J^c \Phi(y) \delta(y-l_s) 
        + \eta^c \psi_x^0(y) \, X \, \Phi^c \right]\nonumber\\[0.2cm]
  & & + H.c.,
\eea
where $m_{\phi}$, $J^c$, and $\eta^c$ are all dimensionful couplings of order the
Planck scale to the appropriate power.  We take $\psi_x^0(y)$ normalized such 
that $\int dy \left| \psi_x^0(y) \right|^2 = 1$
in order to produce canonically normalized kinetic terms.  
The $F$-term equations are
\bea
\left|F_{\phi}(y)\right|&=&\left|(-\partial_y + m_{\phi})\phi^c(y) 
        + J^c \, \delta(y-l_s) \right|\\
\left|F_{\phi^c}(y)\right|&=&\left|(\partial_y + m_{\phi})\phi(y) 
        + \eta^c \, \psi_x^0(y) \, X \right|\\[0.2cm]
\left|F_X^0\right|&=&\left|\eta^c \int dy \, \psi_x^0(y) \, \phi^c(y)\right|.
\eea
For $X$ localized far from $l_s$, the potential is minimized with a non-zero
VEV for $\phi^c$:
\begin{equation}
\langle\phi^c\rangle\simeq-\theta(l_s-y) J^c e^{-m_{\phi}(l_s-y)}.
\end{equation}
If $\psi_x^0(y)$ is a narrow function localized around the point 
$l_x < l_s$, then the $F$-term conditions ($F_i=0$) cannot be 
simultaneously met and thus supersymmetry is broken.  The field $X$ 
now plays the role of the hidden sector singlet\footnote{Again the
full 5d $N=1$ supersymmetry is broken by such couplings.  We assume 
that the theory above $M_*$ accounts for this specific breaking pattern.
A simple high-energy description would be to confine $X$ to a brane
localized in $y$, but we shall leave the discussion more general.} in 
eqs.~(\ref{gauginos}-\ref{scalars}).
As noted below eq.~(\ref{scalars}), provided $X$ is sufficiently
distant from the MSSM matter, it will provide negligible soft masses
to the sfermions.

A $\mu$ term can be generated \cite{SS} by adding the following terms
to the Lagrangian:
\bea
\int d^2 \theta \left[ J \Phi^c(y) \delta(y-l_x) 
        + \lambda_{\mu} \psi_{H_u}^0(y) \psi_{H_d}^0(y) H_u H_d \Phi(y)
        \right].
\eea
Using the notation of Section~\ref{flavor}, including the redefinition
$l_i \rightarrow l_i/\zeta$, and taking $|l_x| \gg 1$ we find
\bea
\mu&=&\int_{l_x}^{\infty} dy \sqrt{\frac{2}{\pi}} \lambda_{\mu} J 
   \sqrt{r_u r_d} \, {\rm exp} \left[ - r_u^2 y^2
   - r_d^2 (y-l_h)^2 - r_{\phi} (y-l_x) \right]\nonumber\\[0.2cm]
&\simeq&\lambda_{\mu} J\sqrt{\frac{2 r_u r_d}{r_u^2+r_d^2}} \,
   {\rm exp} \left[ {r_{\phi} l_x - 
   \frac{r_u^2 r_d^2 l_h^2 + r_d^2 l_h r_{\phi}+(r_{\phi}/2)^2}{r_u^2+r_d^2}}
   \right] \nonumber \\
&\simeq& M_* \, {\rm exp} \left[ {r_{\phi} l_x - 
   \frac{r_u^2 r_d^2 l_h^2 + r_d^2 l_h r_{\phi}+(r_{\phi}/2)^2}{r_u^2+r_d^2}}
   \right],
\label{mu1}
\eea
where $r_{\phi}=m_{\phi}/\zeta$. The first term in the exponential 
dominates and is negative as presumed above.  Taking 
$\psi_x^0(y)=\left(2 \zeta_x^2/\pi\right)^{1/4}
 e^{-\zeta_x^2(y-l_x)^2}$ and $r_x\equiv \zeta_x/\zeta$, we get a
gaugino mass at the compactification scale of
\bea
M_{1/2}&=&\int_{-\infty}^{l_s} dy 
   \left( \frac{2}{\pi} \right)^{1/4}
   \frac{\eta^c J^c}{\sqrt{\zeta}M_*(M_* L)}
   \sqrt{r_x} \, {\rm exp}\left[ -r_x^2(y-l_x)^2 + r_{\phi}(y-l_s)
   \right]\nonumber\\[0.2cm]
&\simeq& (2\pi)^{1/4}\frac{\eta^c J^c}{\sqrt{\zeta r_x}M_*(M_* L)} \, 
   {\rm exp}\left[ {-r_{\phi}(l_x - l_s) 
   + (r_{\phi}/2r_x)^2}\right]\nonumber\\[0.2cm]
&\sim& M_c \, {\rm exp}\left[{r_{\phi}(l_x - l_s) 
   + (r_{\phi}/2r_x)^2} \right].
\label{gauginos1}
\eea
The last lines of eq.s~(\ref{mu1}) and (\ref{gauginos1}) come from taking
$\eta^c \sim J^{2/3} \sim (J^c)^{2/3} \sim \zeta_x \sim M_*$.
By making $l_h \sim l_s + (a \, few)$, the $\mu$ term and $M_{1/2}$
will be the same size.  
Again, a small coupling (required in \cite{SS}) can be generated
by an $\O(1)$ distance (in Planck units).  

Even though $X$ is fixed far from the MSSM matter fields, loop 
contributions of the type in fig.~\ref{loop} can still play an 
important role in the low-energy spectrum if the distance between
$X$ and the matter fields, $\Delta l\equiv (l_i - l_x)$, is a fraction
of the total size $L$ of the extra dimension.  The contribution from
these loops were calculated in \cite{KKS} for the case of $\Delta l=L$.
For $\Delta l \ll L$, the loop integrals are cut off above the mass of
$N_{KK} \sim (L/\Delta l)$ Kaluza-Klein modes.  The integrals require
the 5d gaugino propagator \cite{KKS,AHGS} expressed in position space in
the fifth dimension and in momentum space in the 4 large dimensions,
\bea
P_5(q;x) &=& \left( \frac{2}{L} \right) \sum^{N_{KK}}_n \: 
\frac{ e^{i p_n \, \Delta l} 
\, \left( \gamma^\mu q_\mu + i \gamma_5 p_n \right)}
{q^2 - p_n^2}
\eea
where $p_n = n \pi / L$ is the (quantized)
momentum flowing the the compact dimension.  The sum can be
carried out and results in a simple expression in terms of
hyperbolic functions \cite{AHGS}. Inserting this
propagator in the Feynman diagram shown in fig.~\ref{loop} the
contribution to the scalar masses may be estimated as,
\bea
\label{gauginoloops}
m^2 &\sim& \frac{g_4^2}{16 \pi^2} \, M_{1/2}^2 \, 
\int^\infty_{1} dq \: \frac{q^2 \, {\rm cosh}^2[q (1-x)]}
{{\rm sinh}^2[q] \, {\rm tanh}[q]} .
\eea
where $x = \Delta l / L$, and we have rescaled $q$ by $1/L$ so that the
leading $L$ dependence is absorbed into $M_{1/2}$ and $g_4$, resulting
in the 4d effective quantities appearing in the equation.

We have inserted the entire set of SU(5) gauginos into the loops,
and assumed them all to be massless above the compactification scale.  
This
result should be regarded as an order of magnitude estimate, because
once the theory begins to look 5 dimensional, the gauge coupling will
experience strong running, and though these effects appear in the
calculation formally at higher order, it may be important to resum them
by including the running of the coupling with $q$.  We estimate these
effects to be a factor of a few.

These contributions can in principal be larger than one-loop threshold
corrections and should be included in our high-scale boundary conditions.
Distances of $\Delta l \leq L/4$ could lead to contributions which shift
the LSP from a stau to a neutralino.  What is interesting about these
contributions is that while they are flavor {\it diagonal} they are not
flavor {\it independent} due to the different locations of the generations.
For a wide range in parameter space ($\Delta l$ vs. $L$), these contributions
are below the bounds on additional FCNC and CP violation \cite{GGMS} and
leave a distinct imprint on the spectrum.

\subsubsection{Scenario 2}

For variety, we could move $X$ from where $J$ is to where $J^c$ is (near
the matter fields) and remove $J^c$ altogether.  This corresponds to 
exchanging $l_x$ and $l_s$ above.  In this way, $\phi$'s
VEV is responsible for both supersymmetry breaking and the $\mu$ term.  
In addition,
the coupling $\eta^c X \Phi^c$ must be exchanged for $\eta X \Phi$.  In
principal this would work while giving a non-negligible contribution to
$H_d$ from Planck suppressed operators.  

For successful gaugino mediation, gravity-mediated contributions to 
squark and slepton masses must be small.  Thus, this essentially puts 
a restriction on how close $X$ can be localized to the matter fields.  
For a squark positioned at $l_{i}$, its squared mass $m_i^2$ receives 
a contribution of
\bea
m^2_i &\sim& \frac{F_x^2}{M_*^2} \;
   {\rm exp}\left[{-\frac{2 r_x^2}{r_x^2 + 1}\left[(l_i-l_x)^2 \right]}
\right]
\label{scalardiag}
\eea
while an off-diagonal mass squared $m_{ij}^2$ with squarks localized 
at $l_{i}$ and $l_{j}$ receives a contribution 
\bea
   \frac{F_x^2}{M_*^2} \;
   {\rm exp}\left[
{-\frac{1}{2(r_x^2+1)}\left[(l_i-l_j)^2 
        + 2r_x^2\left((l_i-l_x)^2 + (l_j-l_x)^2 \right) \right]}\right],
\label{scalaroffdiag}
\eea
which is sufficiently suppressed for distances of order a few units and 
$r_x$ of $\O(1)$.  Thus, only $H_d$ receives a non-negligible contribution
to its soft mass for most of parameter space.

However,
the loop contributions described in scenario~1 are too 
large in this scenario because $X$ is so close to the matter fields.  This 
problem can be remedied by altering the shining mechanism.  If instead of 
the coupling $\eta X \Phi$ we add the coupling $\lambda_x X \Phi^2$ then 
we find for $F_X$:
\bea
\left| F_X \right| = \left| \lambda_x \int dy \psi_x^0 (y) \phi(y)^2 \right|,
\eea
thus requiring $l_x$ to have a more intermediate value between $l_h$ and 
$l_s$.  The $\mu$ term in this scenario is the same as in equation 
(\ref{mu1}) if one replaces $l_x$ with $l_s$, while the universal gaugino 
mass is
\bea
M_{1/2}&\simeq&(2\pi)^{1/4}\frac{\lambda_x J^2}{\sqrt{\zeta_x}M_*(M_* L)}
   \, {\rm exp}\left[ {-2r_{\phi}(l_x - l_s) + (r_{\phi}/r_x)^2}\right] 
   \nonumber \\[0.2cm]
   &\sim& M_c \, {\rm exp}\left[{-2r_{\phi}(l_x - l_s)+(r_{\phi}/r_x)^2}
\right] .
\eea

In Section~\ref{pheno} we show that
for small enough $L$, $X$ can give a scalar mass contribution to $H_d$ 
which alters the particle spectrum significantly.  The loop contributions
to scalar masses discussed above still play a small but significant role 
while similar contributions to $A$ terms and $B\mu$ are negligible.
There are no additional CP violating phases coming from soft terms since 
the phases in $M_{1/2}$ and $\mu$ can be rotated away and all of the loop
contributions are to diagonal (and therefore real) soft 
masses\footnote{This however does not solve the strong CP problem}.
\section{Sparticle Spectrum and Phenomenology}
\label{pheno}
\indent \indent
Having introduced the general framework for our model,
we now turn to some specific numbers.
In order to produce a weak scale $\mu$, we fix $l_s$ (for a given
$r_s \sim \O(1)$), such that the scalar VEV of $\Phi$ reaches the
weak scale in the vicinity of the Higgs fields.  Thus, for
a given choice of $r_s$, $l_s$ may be determined such that the
resulting $\mu$ term induces the correct EWSB radiatively
(though in general one must know what the high scale soft
masses are in order to know what value of $\mu$ that is).

\subsection{Sparticle Spectrum}
\indent \indent
Both scenarios for supersymmetry-breaking have similar 
``gaugino-dominated'' boundary conditions, in which the soft
masses for all scalars are smaller than the gaugino masses,
with the possible exception of the down-type Higgs, which receives
the largest gaugino-loop contributions in scenario~1 as well as
sizable supergravity contributions in scenario 2.

This hierarchy in the scalar masses at the GUT scale 
differs in one very important way with respect to
the standard minimal supergravity inspired (SUGRA) models.  The fact that
we have non-universal scalar masses means that there are contributions
to the evolution of the soft masses from the ${\rm U(1)}_Y$ $D$-terms.
These $D$-terms contribute to the beta function of scalar mass $m^2_{i}$,
\bea
\frac{d}{dt}m^2_i &=& Y_i \, \frac{3}{5} \, \frac{g_1^2}{16 \pi^2} 
\; S , \nonumber \\
S &=& \left(
m^2_{H_u} - m^2_{H_d} + 
{\rm Tr} \left[ m^2_{Q} - 2 m^2_{u} + m^2_{d} -
m^2_{L} + m^2_{e} \right] \right) , 
\eea
where $g_1$ is the $U(1)_Y$ coupling, normalized appropriately
for SU(5) unification, $Y_i$ is (standard) hypercharge for scalar $i$,
and the trace is over the three families.  The quantity $S$ is an RG
invariant, and thus vanishes at all scales if all scalar masses are
equal at some scale.  Thus it contributes nothing to 
the evolution of SUGRA scalar masses.
In our model, this term will affect all of the scalar masses, with the
effect being the most dramatic for the sleptons, which have large
hypercharges and receive no contributions from the strong coupling.
This results in a neutralino LSP in a large region of parameter space.  

\FIGURE[t]{
\epsfysize=4.0in
\centerline{\epsfbox{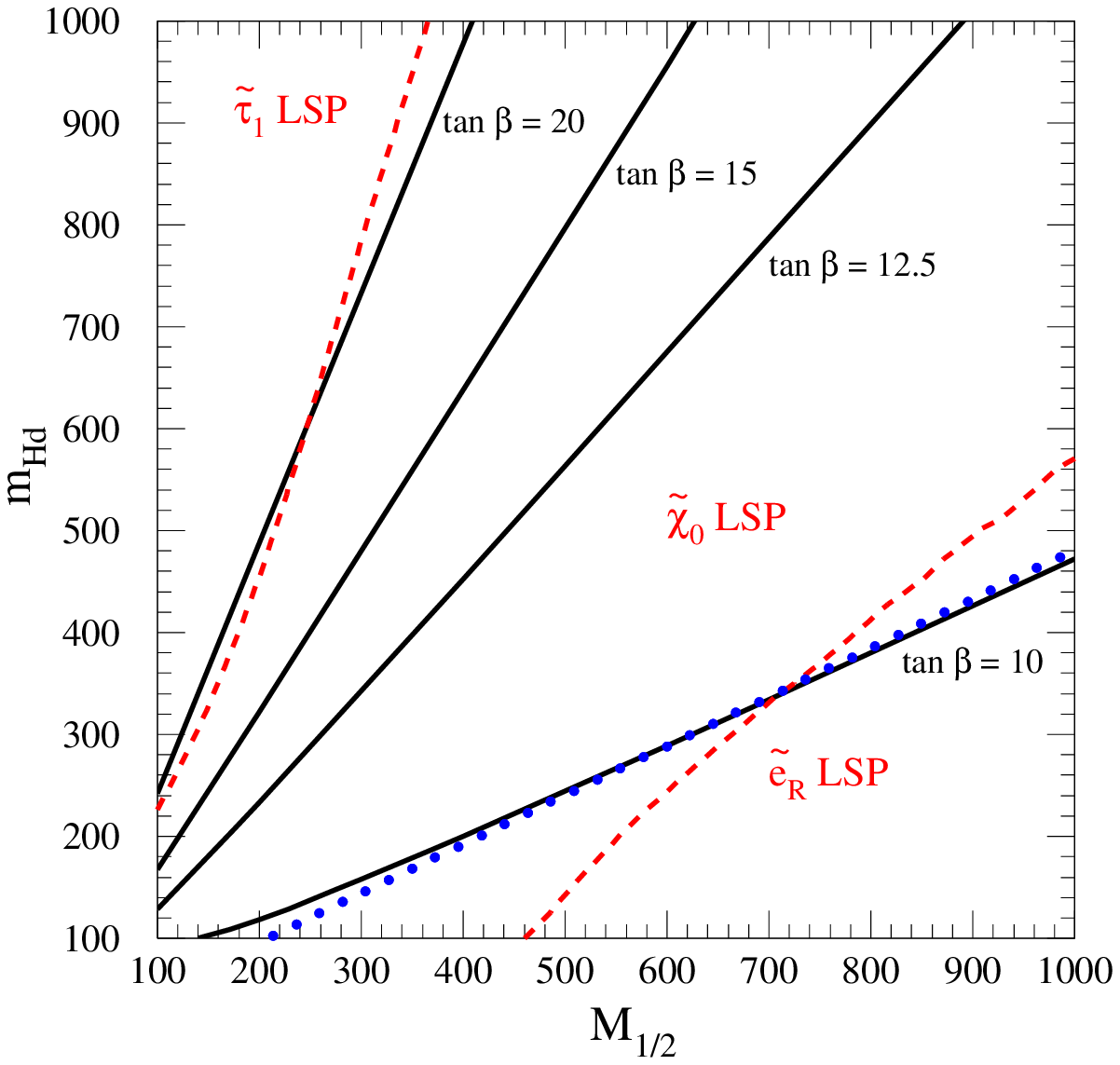}}
\caption{The solid curves are the
contours of $m_{H_d}$ for a choice of $M_{1/2}$ (both at the
GUT scale) and $\tan \beta$.
The dashed curves demark regions where $\widetilde{\tau}_1$,
$\widetilde{e}_R$, and $\widetilde{\chi}^0_1$ are the LSP.
The dotted curve marks the value of $m_{H_d}$ as a function of
$M_{1/2}$ for the example model of scenario 1.}
\label{m12mhfig}}

The condition that
$B\mu=0$ at the high scale provides us with a particular moderate
to high value of $\tan \beta$ \cite{hightanb} for each choice
of $M_{1/2}$ and $m_{H_d}$.  In fig.~\ref{m12mhfig} we show the
contours of constant $\tan \beta$ in the plane of $M_{1/2}$ and
$m_{H_d}$, including the gaugino-loop contributions
given in eq.~(\ref{smallloops}).
It is remarkable that both the mechanism that generates the 4d
Yukawa couplings of Section~\ref{flavor} and the scenarios for
supersymmetry breaking favor the same intermediate to large
values of $\tan \beta$.

The resulting
LSP in the plane of $M_{1/2}$ and $m_{H_d}$ 
is indicated in fig.~\ref{m12mhfig} by the dashed curves\footnote{This
figure has included scalar masses at the GUT scale introduced
in eq.~(\ref{smallloops}) below.  These contributions to the
scalar masses at $M_c$ do not significantly affect the contours of
$\tan \beta$, though it does generally alter the LSP curves}.  
From this figure,
we see that if $M_{1/2}$ is much larger than $m_{H_d}$, 
we arrive at a $\tilde{e}_R$ LSP because the large $M_{1/2}$ results in
a relatively large neutralino mass, while the selectrons get very little
contributions from gaugino loops and the $D$-term contribution to the
evolution is small for small $m_{H_d}$.
On the other hand,
if $m_{H_d}$ is much larger than $M_{1/2}$, the $D$-term contribution
to the slepton masses overpowers the standard gaugino contribution.
This
results in a $\widetilde{\tau}_1$ LSP (which is mostly
$\widetilde{\tau}_L$) or a negative $m^2$ for
$\widetilde{\tau}_1$ which spontaneously breaks the electromagnetic 
symmetry (for consequences of a generic $D$-term, see \cite{LMSSM}).
However, for $m_{H_d} \sim M_{1/2}$ the $D$-term contribution is enough
to raise the $\widetilde{e}_R$ mass above the lightest 
neutralino mass, and does not push the $\widetilde{\tau}_1$ below it,
resulting in a stable $\widetilde{\chi}_0$ LSP which provides
a suitable dark matter candidate, and results in missing transverse energy
signatures at a hadron collider.

\subsection{Scenario 1}
\indent \indent
In this scenario, the gaugino masses arise from a superpotential
coupling $X \, \Phi$ which induces a VEV in $F_X$.
$X$ is located a large distance
away from the matter fields.  Thus, all supergravity contributions to
scalar masses are zero because of the small overlap between the
distant $X$ field and the matter fields.  However, there are
contributions from the gaugino-loops in (\ref{gauginoloops})
which contribute to all of the scalar masses, and thus are all
proportional to $M_{1/2}$.  Thus, all of the boundary conditions
are effectively specified by $M_{1/2}$ and the position of $X$.

As a particular example, for $|l_x - l_h| / L \sim 0.4$ one 
obtains\footnote{Though eq.(\ref{gauginoloops}) is only an estimate
of the loop contributions, it preserves the percentage difference between
scalar masses reasonably accurately.  Thus, we present figures to two
significant digits to better illustrate the mass splittings.},
\begin{alignat}{3}
\label{smallloops}
m_{\tilde{f}_1}^2 &= 0.014 \, M_{1/2}^2 , &\qquad
m_{H_d}^2 &= 0.23 \, M_{1/2}^2 , \nonumber \\
m_{\tilde{f}_2}^2 &= 0.019 \, M_{1/2}^2 , &\qquad
m_{H_u}^2 &= 0.61 \, M_{1/2}^2 , \nonumber \\
m_{\tilde{f}_3}^2 &= 0.049 \, M_{1/2}^2 . &\qquad &
\end{alignat}
Thus, our soft terms for squarks
are dominated by the gaugino-mediated contributions, and are relatively
flavor-blind.  The off-diagonal squark matrix entries are then
induced by the CKM rotation from interaction to mass eigenbasis
and can be estimated for the first two families as,
\bea
\frac{\widetilde{m}_d^2 - \widetilde{m}_s^2}{\widetilde{m}_s^2} \, 
V_{us} \sim 4 \times 10^{-4}
\eea
This is small enough to avoid the supersymmetry flavor 
and CP problems generally
associated with off-diagonal squark masses.
The dotted line in fig.~\ref{m12mhfig} shows the value of $m_{H_d}$ for
this model.  As can be seen, this corresponds to $\tan \beta \sim 10$
and has a neutralino LSP for $M_{1/2} < 700$ GeV.

\subsection{Scenario 2}
\indent \indent
In this scenario
gaugino masses are determined by the singlet $X$ which develops
an auxiliary VEV through superpotential coupling $X \, \Phi^2$.
$M_{1/2}$ may be fixed by localizing $X$ a suitable 
distance between the
Higgses and the source for $\Phi$ so that $F_X / (M^2_* L)$ is
a weak scale gaugino mass (we assume that
$L \sim l_s$, or in other words, that the source for $\Phi$ is
localized on one end of the compact dimension, and the MSSM
matter and Higgs live roughly at the other end). 
The $X$ field must be localized far enough from the fermions to
avoid dangerous supergravity-mediated flavor mixing in the soft masses.  
We can proceed by choosing $r_s$ and $l_s$ to get a particular
weak scale $\mu$ and localize $X$ so that an appropriate $M_{1/2}$
results.
For some ``typical'' numbers $r_s = 2$ (and $l_s = -16.5$ so that
$\mu \sim 450$ GeV), these two constraints require that 
$X$ be localized in the region $l_x \sim -4.7$ for
$r_x \sim 0.5$.  This results in
a gaugino mass of $M_{1/2} \sim 300$, which is
the right order of magnitude to provide the correct EWSB
given the value of $\mu$.  

Computing the supergravity contributions
to the soft masses, we find that all of the squark and the
up-type Higgs receive negligible contributions because they
are localized too far away from $X$.  On the other hand,
the down-type Higgs receives a substantial contribution to its soft mass
of about $m_{H_d}^2 \sim ({\rm 250~GeV})^2$ because it lies relatively 
close to $X$.  The 5d loop contributions are also generically small
because they are loop suppressed, and have been chosen for this example
to be the same as those presented in (\ref{smallloops}), and are thus
once again safe from the point of view of flavor violation.

By performing a brute force
scan through the ``reasonable'' range of $r_s$ and
$r_x$ from 0.5 to 2, we find that this is a generic prediction of the model;
we are able to obtain any $M_{1/2}$ and 
$m_{H_d}$ between 100 GeV and 1~TeV, as well
as the corresponding value of $\mu$ at the high scale required to
induce EWSB from these boundary conditions.  
The supergravity contributions to
other soft masses are always so small as to be negligible,
and the gaugino-loop contributions to squark masses are
a small factor times $M_{1/2}^2$.

Having determined that our model easily can realize the plane of
$M_{1/2}$ and $m_{H_d}$ from the regions of 100 GeV to 1 TeV,
and found that we can accommodate the necessary $\mu$ for
EWSB resulting $\tan \beta$, we can switch our discussion from the
underlying model parameters $r_s$, $l_s$, $r_x$, and $l_x$ and
discuss the resulting theory defined by a particular choice of
$M_{1/2}$ and $m_{H_d}$ at scale $M_c$ (which we will take for
simplicity to be the GUT scale, 
$M_{GUT} \sim 2 \times 10^{16}$~GeV).  
Thus, we may completely specify the
supersymmetry breaking parameters of scenario 2 by the two free parameters,
\bea
M_{1/2}, \: m_{H_d}
\eea
with $\mu$ fixed for radiative EWSB and $\tan \beta$ determined by
the condition that $B \mu = 0$ at $M_c \sim M_{GUT}$.
As shown in fig.~\ref{m12mhfig},
this results in a particular value of $\tan \beta$ determined by the
boundary conditions $B \mu =0$ at $M_c$, while  $\mu$ is determined at the
weak scale from the observed mass of the $Z$ boson, and depends
very strongly on the choice of $M_{1/2}$ and rather weakly on the
choice of $m_{H_d}$.

\subsection{Collider Signatures}
\indent \indent
The resulting weak scale phenomenologies for both 
supersymmetry breaking scenarios 
discussed above are similar, and thus we briefly discuss both together
here.  As in all gaugino-dominated models, the sparticle spectrum
is mostly dependent on 
the choice of $M_{1/2}$.  For example, for $M_{1/2} \sim m_{H_d} \sim 200$
(so $\tan \beta \sim 12$), the resulting spectrum has squarks with masses
on the order of 430~GeV, charged and neutral
sleptons with masses between 80 and 150~GeV, gluinos with mass around
500~GeV, and weak charginos and neutralinos 
with masses between 75 and 320~GeV, with
the lighter states being dominantly gaugino-like.  A
number of these particles would be accessible at Run~II of the 
Tevatron, with a variety of signals (all characterized by the missing
transverse energy from the $\widetilde{\chi}_0$ LSP).  
Larger
values of $M_{1/2}$ will result in a heavier sparticle spectrum,
with the qualitative features unchanged; the gluino will tend to be the
heaviest superparticle, with the squarks somewhat lighter, followed by
sleptons and weak gauginos.  Once the superpartner masses become
heavier than a few hundred GeV there is insufficient energy to produce
them at the Tevatron.  However, the CERN 
Large Hadron Collider (LHC) will probe $M_{1/2}$ up to roughly one TeV.
Again, this will be realized through a variety of signals, one particular
interesting example of which is the ``tri-lepton'' signal coming from the
leptonic decay of the lighter gauginos into the neutralino LSP
\cite{trilepton}.

As in all models which can be described at low energies by the MSSM,
there are three neutral Higgs bosons (two CP even and one CP odd)
and a pair of charged Higgs scalars.  The heavy CP even, CP odd,
and charged Higgs bosons typically have masses that are a few times
larger than $M_{1/2}$ and thus the model exhibits the Higgs
``decoupling'' limit in which the lightest Higgs boson has 
approximately SM couplings.
For the moderate values of $\tan \beta$ realized by the model,
the interesting signals of the pseudo-scalar Higgs produced
in association with $b$ quarks may be observed at Tevatron or LHC
\cite{bhiggs}.
The lightest Higgs boson
typically has a mass in the range of 110 to 120 GeV, much of which
will be probed by LEP \cite{lephiggs}, 
with higher masses typically accessible to the Tevatron and/or LHC
\cite{higgs}.
\section{Conclusions}
\label{conclusions}
\indent \indent
We have presented a model in which all masses below the GUT scale are
generated by localizing fields in a small extra dimension.  All 
dimensionless couplings are of order one, all dimensionful couplings are
of order the Planck scale and all distances are of order the Planck
length, save the size of the extra dimension which is of order
the inverse GUT scale.  The quark flavor portion of the model
manages to beautifully realize the CKM mixings and CP-violating
phase observed in nature, as well the observed quark masses.  It is
completely independent of the nature of supersymmetry breaking, 
and thus can be taken as a generic picture of how quark masses
might arise from a small extra dimension in a supersymmetric context.

We have explored two different pictures for how the extra dimension
might play a role in supersymmetry breaking, both of which are variations of
gaugino mediation.  While successfully incorporating attractive
features such as radiative EWSB and the possibility of a 
neutral LSP suitable as a dark
matter candidate, they result in distinctive boundary conditions for
soft masses at the GUT scale, and thus result in interesting
relations among superparticle masses not seen in other models.
The separation of the families leaves an imprint on the boundary
conditions, producing a striking scalar spectrum distinguishable
from other predictive models of supersymmetry breaking.

We have employed a bottom-up approach in constructing the model,
taking for granted details related to the spontaneous breaking
of the GUT symmetry to the SM gauge group, specific details
concerning the dynamical localization of the fermions, and the
breakdown of the full $N=1$ supersymmetry in the 5d theory to the $N=1$
in the 4d theory.  It would be interesting to pursue these last
two technical details with more rigor.
Further, it would be interesting to construct a full GUT theory,
to explore the possibility that GUT physics
stabilizes the size of the extra dimension at the GUT scale.
Further, the details of the GUT breaking must address the lepton
masses observed in nature.  In fact,
our model contains (at least) two gauge singlet fields which 
might play the role of the right-handed
neutrino.

In general, we find that a small extra dimension allows for a vast
array of new possibilities to solve old problems.  While extra dimensions
of this size would not be accessible at any collider in the near future,
their indirect effects on low energy phenomena can be dramatic.

\section{Acknowledgements}
It is a pleasure to acknowledge discussions with H.--C. Cheng,
H.--J. He, 
M. Schmaltz, and C.--P. Yuan.  C.E.M. Wagner has provided an
invaluable source of discussion, comments, questions, and encouragement.
Work at Argonne National Lab 
is supported in part by the DOE under contract W-31-109-ENG-38.  
DEK is also supported in part by the DOE under contract 
DE-FG02-90ER40560.  
\appendix
\section{Localizing a Chiral Superfield in an Extra Dimension}
\label{local}
\indent \indent
In this appendix we show how a chiral superfield may be localized in
an extra dimension with an exponential profile.  Our starting point is
the (on-shell) action for an $N=1$ chiral superfield in 5 space-time
dimensions,
\bea
\label{act1}
S &=& \int d^4x \; dy \left\{ \int d^4\theta \; 
\left(  {\Psi}^\dagger  {\Psi} 
+ { {\Psi}_c}^\dagger { {\Psi}}_c \right) 
\right. \nonumber \\
& & \left.
+ \left( \int d^2\theta \; 
i \;  {\Psi}_c [ \partial_y + M(y) ]  {\Psi}
+ H.c. \right) \right\} ,
\eea
where $ {\Psi}$ and $ {\Psi}_c$ are the 
left-chiral and charge-conjugated right-chiral $N=1$ 4d
superfield components of the single $N=1$ 
5d chiral superfield $ {\Psi}$.
$\partial_y = \partial / \partial y$ denotes the partial 
derivative with respect to the fifth dimension. 
This formulation of the action is convenient because it is written in
the usual $N=1$ superfield language and thus is manifestly $N=1$
supersymmetric.
We have written the mass parameter $M(y)$ explicitly as a function of the
fifth dimension.  This could be realized, for example, by 
appropriately coupling an additional chiral superfield to 
$ {\Psi}_c \,  {\Psi}$ and
including dynamics which give its scalar component a VEV.

In terms of component fields, eq.~(\ref{act1}) may be written,
\bea
\label{com1}
S &=& \int d^4x \; dy \left\{ \, F_L^* \, F_L \: + \: F_R^* \, F_R \: 
+ \: \partial_\mu  {\phi}_L^* \, \partial^\mu  {\phi}_L \:
+ \: \partial_\mu  {\phi}_R^* \, \partial^\mu  {\phi}_R \right. 
\nonumber \\
& & \left. \:
+ \, i \, \overline{\Psi} \, 
  \left[ \gamma^\mu \, \partial_\mu - \gamma_5 \, \partial_y
  - M(y) \right] \Psi 
\right. \nonumber \\[0.2cm]
& & \left. \:
+ \, i \left(  \: F_R^* \, [ \partial_y + M(y) ] \,  {\phi}_L
          \: + \:  {\phi}_R^* \, [ \partial_y + M(y) ] \, F_L
\right. \right. \nonumber \\[0.2cm]
& & \left. \left. \:
          + \: F_L^* \, [ \partial_5 - M(y) ] \,  {\phi}_R
       \: + \:  {\phi}_L^* \, [ \partial_y - M(y) ] \, F_R \right)
\right\} \: .
\eea
where $ {\phi}_L, P_L \Psi, F_L$ ($ {\phi}_R^*, P_R \Psi, F_R^*$)
are the scalar, spinor, and $F$ component fields of $ {\Psi}$ 
($ {\Psi}_c$), respectively.  Expanding these fields in terms of a 
complete orthonormal basis of scalar functions of $y$,
\begin{alignat}{3}
\label{modes1}
P_L \Psi(x, y) &= \sum_{n} \Psi_L^n(x) \, b^n(y), &\qquad
P_R \Psi(x, y) &= \sum_{n} \Psi_R^n(x) \, f^n(y), 
\nonumber \\
 {\phi}_L(x,y) &= \sum_{n}  {\phi}^n_L(x) \, B^n(y), &\qquad
 {\phi}_R(x,y) &= \sum_{n}  {\phi}^n_R(x) \, F^n(y),
\nonumber \\
F_L(x,x5) &= \sum_{n} F_L^n(x) \, {\cal B}^n(y), &\qquad
F_R(x,x5) &= \sum_{n} F_R^n(x) \, {\cal F}^n(y),
\end{alignat}
we find that the action eq.~(\ref{com1}) 
simplifies considerably if one requires,
\bea
\label{lineq}
\left[ \, \partial_y + M(y) \, \right] \left\{ 
\begin{array}{c} b^n(y) \\ B^n(y) \\ {\cal B}^n(y) \end{array} 
\right\} = \lambda_n \, \left\{
\begin{array}{c} f^n(y) \\ {\cal F}^n(y) \\ F^n(y) \end{array} 
\right\} , \nonumber \\[0.2cm]
\left[ \, \partial_y - M(y) \, \right] \left\{ 
\begin{array}{c} f^n(y) \\ F^n(y) \\ {\cal F}^n(y) \end{array} 
\right\} = -{\lambda_n}^* \, \left\{
\begin{array}{c} b^n(y) \\ {\cal B}^n(y) \\ B^n(y) \end{array} 
\right\} ,
\eea
indicating that [$b^n(y)$, $f^n(y)$], [$B^n(y)$, ${\cal F}^n(y)$],
and [${\cal B}^n(y)$, $F^n(y)$] are ``bosonic'' and ``fermionic''
pairs of solutions to a SUSY Quantum Mechanics problem with
$Q = [\partial_y + M(y)]$ \cite{kaplanf}.  These first order
differential equations may be combined to give second order equations
for each individual function,
\bea
\label{quadeq}
\left[ \, -\partial^2_y + M^2(y) - (\partial_y M(y)) \, \right] \left\{ 
\begin{array}{c} b^n(y) \\ B^n(y) \\ {\cal B}^n(y) \end{array} 
\right\} = |\lambda_n|^2 \, \left\{
\begin{array}{c} b^n(y) \\ B^n(y) \\ {\cal B}^n(y) \end{array} 
\right\} , \nonumber \\[0.2cm]
\left[ \, -\partial^2_y + M^2(y) + (\partial_y M(y)) \, \right] \left\{ 
\begin{array}{c} f^n(y) \\ F^n(y) \\ {\cal F}^n(y) \end{array} 
\right\} = |\lambda_n|^2 \, \left\{
\begin{array}{c} f^n(y) \\ F^n(y) \\ {\cal F}^n(y) \end{array} 
\right\} .
\eea
From this result, we see that the same differential equation
determines $b^n(y)$, $B^n(y)$, and ${\cal B}^n(y)$,
and another determines $f^n(y)$, $F^n(y)$, and ${\cal F}^n(y)$,
and thus there are only two independent sets of functions relevant
to the four dimensional effective theory description (instead of six).
This is certainly not surprising; it is an indication that 4d $N=1$ supersymmetry
remains unbroken.  The important point for
our purposes is that each component field for a given KK mode of the
chiral superfield acquires the same profile in the extra dimension.
Putting this expansion into eq.~(\ref{com1}) and carrying out the
$dy$ integration, we arrive at,
\bea
\label{finalact}
S &=& \sum_n \int d^4x \: {F^n_L}^* \, F^n_L \: 
      + \: {F^n_R}^* \, F^n_R \: 
+ \: \partial_\mu  {\phi}_L^{n *} \, 
     \partial^\mu  {\phi}_L^n \:
+ \: \partial_\mu  {\phi}_R^{n *} \, 
     \partial^\mu  {\phi}_R^n \\
& & \,
+ \, i \, {\overline{\Psi^n}} \, 
  \left[ \gamma^\mu \, \partial_\mu + \lambda_n \right] \Psi^n 
+ \, i \lambda_n \left(  
            {F^n_R}^* \,  {\phi}^n_L
          +  {\phi}_R^{n *} \, F^n_L
          + {F^n_L}^* \, { {\phi}^n}_R
          +  {\phi}_L^{n *} \, F^n_R \right) 
\nonumber \\
 &=&  \sum_n \int d^4x \left\{ \int d^4\theta \; 
\left(  {\Psi}^{n \dagger}  {\Psi}^n 
+ { {\Psi}_c}^{n \dagger} { {\Psi}^n}_c \right) 
+ \left( \int d^2\theta \, 
i \lambda_n \,  {\Psi}^n_c \,  {\Psi}^n
+ H.c. \right) \right\}
\nonumber .
\eea
which is a free supersymmetric theory for an infinite tower
of massive left- plus right-chiral multiplets as well as some number
of left- and right-chiral zero mass modes (whose number
need not be equal \cite{witten}).  Introducing interactions to this
theory does not change the end result for the kinetic terms, though
generically interactions will be induced among all of the various
modes, and with varied coupling strengths proportional to the overlap
of the profiles of all participating modes in the extra dimension.

We will be considering Gaussian profiles for the zero modes,
which can be considered ``generic'' in the sense that
given any mass function which crosses zero at some point in $y$,
for a small region about that point, the function may be
approximated as linear, $M(y) \approx 2 \zeta^2 y$,
with $2 \zeta^2$ the slope at the crossing point.
An example of such a profile is a domain wall arising from a kink soliton.
For this choice of $M(y)$, eq.~(\ref{quadeq}) looks like the
Schr\"{o}dinger equation for a harmonic oscillator with frequency
$2 \zeta^2$ and energy shifted by $-\zeta^2$ for the left-chiral
and $+\zeta^2$ for the right-chiral modes.  Thus, 
for $\zeta^2 > 0$ there is a single
left-chiral zero mode with profile,
\bea
\label{hoscgs}
b^0(y) &=& \left( \frac{2 \zeta^2}{\pi} \right)^{(\frac{1}{4})} \:
             e^{i \, \varphi} \:
             e^{-\zeta^2 y^2} \: ,
\eea
and the higher mode $b^n(y)$ are given by the familiar product of
exponentials and Hermite polynomials.  The $n \geq 1$ $f^n(y)$
functions may be obtained from eq.~(\ref{lineq}).  For $\zeta^2 < 0$
there is a single right-chiral zero mode.  Thus, any crossing of zero
in $M(y)$ will generally produce a zero mode localized
around the zero crossing with Gaussian fall-off 
(the width of the Gaussian being determined by the slope of
$M(y)$ at the crossing point), provided
the deviations from linearity occur sufficiently far away
from the crossing point.  

eq.~(\ref{hoscgs}) explicitly includes a phase $\varphi$.
Such a phase is undetermined by our choice of $M(y)$, and does not
contribute to the kinetic terms derived in eq.~(\ref{finalact}).
However, it may play a role in the interaction terms for the 4 dimensional
effective theory, where careful analysis is required to determine
which phases are physical, and which may be rotated away by an
appropriate redefinition of fields.
\def\pl#1#2#3{{\it Phys. Lett. }{\bf B#1~}(#2)~#3}
\def\zp#1#2#3{{\it Z. Phys. }{\bf C#1~}(#2)~#3}
\def\prl#1#2#3{{\it Phys. Rev. Lett. }{\bf #1~}(#2)~#3}
\def\rmp#1#2#3{{\it Rev. Mod. Phys. }{\bf #1~}(#2)~#3}
\def\prep#1#2#3{{\it Phys. Rep. }{\bf #1~}(#2)~#3}
\def\pr#1#2#3{{\it Phys. Rev. }{\bf D#1~}(#2)~#3}
\def\np#1#2#3{{\it Nucl. Phys. }{\bf B#1~}(#2)~#3}
\def\xxx#1{{\tt [#1]}}

\end{document}